# Observation of micro-macro entanglement of light

A. I. Lvovsky[a,b], R. Ghobadi[a,c], A. Chandra[a], A. S. Prasad[a], C. Simon[a]

**Schrödinger's famous thought experiment[1] involves a (macroscopic) cat whose quantum state becomes entangled with that of a (microscopic) decaying nucleus. The creation of such micro-macro entanglement is currently being pursued in several fields, including atomic ensembles[2], superconducting circuits[3], electro-mechanical[4] and opto-mechanical[5,6] systems. For purely optical systems, there have been several proposals[7,8,9] to create micro-macro entanglement by greatly amplifying one half of an initial microscopic entangled state of light, but experimental attempts[10] have so far been inconclusive[11]. Here we experimentally demonstrate micro-macro entanglement of light. The macro system involves over a hundred million photons, while the micro system is at the single-photon level. We show that microscopic differences (in field quadrature measurements) on one side are correlated with macroscopic differences (in the photon number statistics) on the other side. Further, we demonstrate entanglement by bringing the macroscopic state back to the single-photon level and performing full quantum state tomography of the final state. Our results show that it is possible to create and demonstrate micro-macro entanglement for unexpectedly large photon numbers. Schrödinger's thought experiment was originally intended to convey the absurdity of applying quantum mechanics to macroscopic objects. Today many quantum physicists believe that quantum principles in fact apply on all scales. By combining the present approach with other (e.g. mechanical) systems, or by applying its basic ideas in different contexts, it may be possible to bring quantum effects ever closer to our everyday experience.**

"Schrödinger cats" are notoriously difficult to generate and observe because even the minutest interactions of the system with the environment entangle the two, thereby decohering the superposition. In the optical domain, decoherence is mainly due to losses associated with absorption and spurious reflection at interfaces. However, certain optical states exhibit surprising robustness with respect to such losses, and can be truly macroscopic, yet maintain properties of a quantum superposition.

---

[a] Institute for Quantum Science and Technology and Department of Physics and Astronomy, University of Calgary, Calgary T2N1N4, Alberta, Canada
[b] Russian Quantum Center, 100 Novaya St., Skolkovo, Moscow 143025, Russia
[c] Department of Physics, Sharif University of Technology, Tehran, Iran

For example, Ref. 7 proposed to create micro-macro entanglement of light by starting with a polarization-entangled photon pair and greatly amplifying one of the photons. Ref. 10 claimed to have demonstrated micro-macro entanglement involving $10^4$ photons on the macro side following this approach. However, these results were later shown to be inconclusive by pointing out that, for the measurement procedure used in Ref. 10, equivalent results could be obtained with a separable state[12,13]. It was subsequently understood[11] that, although the state of Ref. 10 is robust to losses, it is very difficult to detect micro-macro entanglement via direct measurements (such as photon counting) on the macroscopic state, because the relevant measurements need to have extremely high resolution. This issue may be resolved by bringing the macroscopic state back to the single-photon level by inverting the amplification operation[14].

The type of amplification considered in the above references was based on optical non-linearities (squeezing). A significantly simpler approach is to use the phase space displacement operation to render the state in one or both channels macroscopic[8]. One can start with the delocalized single-photon state[15]

$$|\Psi\rangle = \frac{1}{\sqrt{2}}\left(|0\rangle_A \otimes |1\rangle_B + |1\rangle_A \otimes |0\rangle_B\right), \quad (1)$$

where $A$ and $B$ refer to fictitious observers Alice and Bob, and apply the phase-space displacement operator $\hat{D}(\alpha) = e^{\alpha \hat{a}^\dagger - \alpha^* \hat{a}}$ to Bob's mode to obtain

$$|\Psi_D\rangle = \frac{1}{\sqrt{2}}\left(|0\rangle_A \otimes \hat{D}(\alpha)|1\rangle_B + |1\rangle_A \otimes \hat{D}(\alpha)|0\rangle_B\right), \quad (2)$$

where $\alpha$ is the macroscopic displacement vector [Fig. 1(a)]. The resulting state is an attractive candidate for the observation of micro-macro entanglement. Surprisingly, even though the displaced single-photon and vacuum states are close in phase space and in mean photon numbers ($\langle N \rangle \approx \alpha^2$), they are macroscopically different in the photon number variance. This property makes the state (2) a macroscopic quantum superposition according to the most basic definition[16], namely a superposition of two states with macroscopically different values for a physical observable. The necessary phase-space displacement is easy to implement in the lab: this is done by overlapping the target state with a strong coherent state on a highly asymmetric beam splitter[17,18].

State (2) is only weakly sensitive to losses[8], which is very advantageous from the point of view of experimental implementation. In contrast, its sensitivity to phase noise increases with the size of the displacement[8], making it essential to implement a highly phase-stable setup. This increasing sensitivity to a decoherence mechanism can be seen as an additional argument for the macroscopic character of the superposition (2)[19].

Finally, one can easily verify the entangled nature of state (2). To that end, one can undo the displacement in Bob's channel by applying operator $\hat{D}(-\alpha)$ to it, bringing state (2) back to microscopic form[8,14] (1), and characterizing it by homodyne tomography[20].

In the present work, we implement state (2) and test it for the two salient features of Schrödinger's cat: macroscopicity and entanglement. First, we verify that, by changing the conditions of a microscopic measurement on Alice's channel and conditioning on specific results of that measurement, we obtain states with macroscopically distinct photon number statistics in Bob's channel. Second, we perform homodyne tomography on the "undisplaced" state and verify that the entanglement has been preserved through the displacement and undisplacement operations.

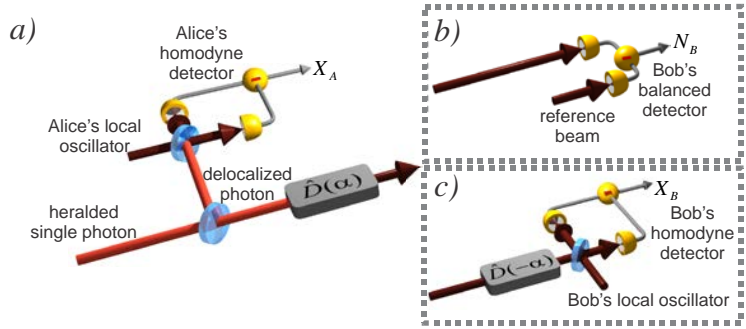

Figure 1: Scheme of the experiment. Panel (a) shows the preparation of the micro-macro entangled state (2) and Alice's measurement of the field quadratures in the microscopic portion of the state. Panels (b) and (c) show the two options for Bob's measurement of the macroscopic portion of the state: energy measurement to verify macroscopicity (b) and "undisplacement" followed by the quadrature measurement to verify entanglement (c). Beams of bright red color correspond to microscopic optical states; dark red to macroscopic.

The principal scheme of the first part of the experiment is shown in Fig. 1 (a,b). A heralded single photon from a parametric down-conversion setup propagates through a symmetric beam splitter to generate the nonlocal single-photon state. We perform a phase-space quadrature measurement in Alice's mode by means of a balanced homodyne detector[21]. At the same time, Bob's mode is subjected to phase-space displacement with $\alpha^2 \sim 1.6 \times 10^8$ photons, after which its photon number $N_B$ is measured.

These energy measurements exhibit macroscopic quantum fluctuations whose statistics are correlated with Alice's measurements of the field quadrature (Fig. 2). This can be qualitatively understood as follows. Alice's measurement of the position observable $X_A$ collapses the entanglement, projecting Bob's mode onto state[22]

$$|\psi_B\rangle = \frac{1}{\sqrt{2}}\left(\psi_0(X_A)\hat{D}(\alpha)|1\rangle_B + \psi_1(X_A)\hat{D}(\alpha)|0\rangle_B\right), \qquad (3)$$

where $\psi_{0,1}(X)$ are the wavefunctions of the zero- and one-photon states in the position basis. If $X_A$ is close to zero, we have $|\psi_1(X_A)| \ll \psi_0(X_A)$, so the state in Bob's channel is close to $\hat{D}(\alpha)|1\rangle$ and its photon number noise variance is about $\langle \Delta N^2 \rangle \sim 3\alpha^2$. On the other hand, in case Alice observes a high quadrature value $X_A \gg 1$, Bob's mode is projected onto a state close to $\hat{D}(\alpha)|0\rangle$ so $\langle \Delta N^2 \rangle \sim \alpha^2$. In this way, projecting onto different values of a microscopic observable at Alice's end leads to macroscopically different photon number statistics at Bob's.

Although ideally the ratio between the photon number variances in these two situations equals 3, in our experiment this number is reduced to about 1.35, primarily due to two effects. First, the observed data are influenced by the imperfection in the preparation of the single-photon state and linear losses, which manifest themselves as admixture of the vacuum state $|0\rangle_A \otimes |0\rangle_B$ to the ideal state (1) [23]. In this work, the vacuum fraction is $1-\eta = 0.46$. Second, we measure the photon number by means of a balanced photodetector[24]. Bob's mode is incident onto the sensitive area of one of its photodiodes while the other photodiode is illuminated by a reference laser pulse of exactly the same mean energy. The subtraction signal is then proportional to $N_B - N_R$, where $N_R$ is the number of photons in the reference pulse. This technique is necessary because the photon number fluctuations of the displaced field are on the scale of α, whereas its mean is much higher: $\langle N_B \rangle \approx \alpha^2$. Subtraction of the reference pulse permits elimination of this background along with its classical noise. As a trade-off, it leads to addition of the shot noise $\langle \Delta N_R^2 \rangle = \alpha^2$ to the signal, thereby reducing the observed ratio of the photon number variances.

The experimental results (Fig. 2) exhibit different behavior dependent on the relative phase of Alice's quadrature measurement and Bob's displacement. If the two are the same, we observe that not only the variance, but also the mean of the photon number observed in Bob's channel are correlated with Alice's results. On the other hand, if the phases are orthogonal, the mean photon number is almost constant. Therefore, by choosing which quadrature to measure, Alice can influence the state prepared in Bob's channel. This is a consequence of the entangled nature of state (2); similar phenomena have been observed in discrete [25], continuous [26] and hybrid[22] systems, but not yet on a macroscopic level. In particular, this behavior explicitly shows absence of decoherence of the two terms in Eq. (2). If such decoherence were present, we would observe no dependence on Alice's choice of quadratures.

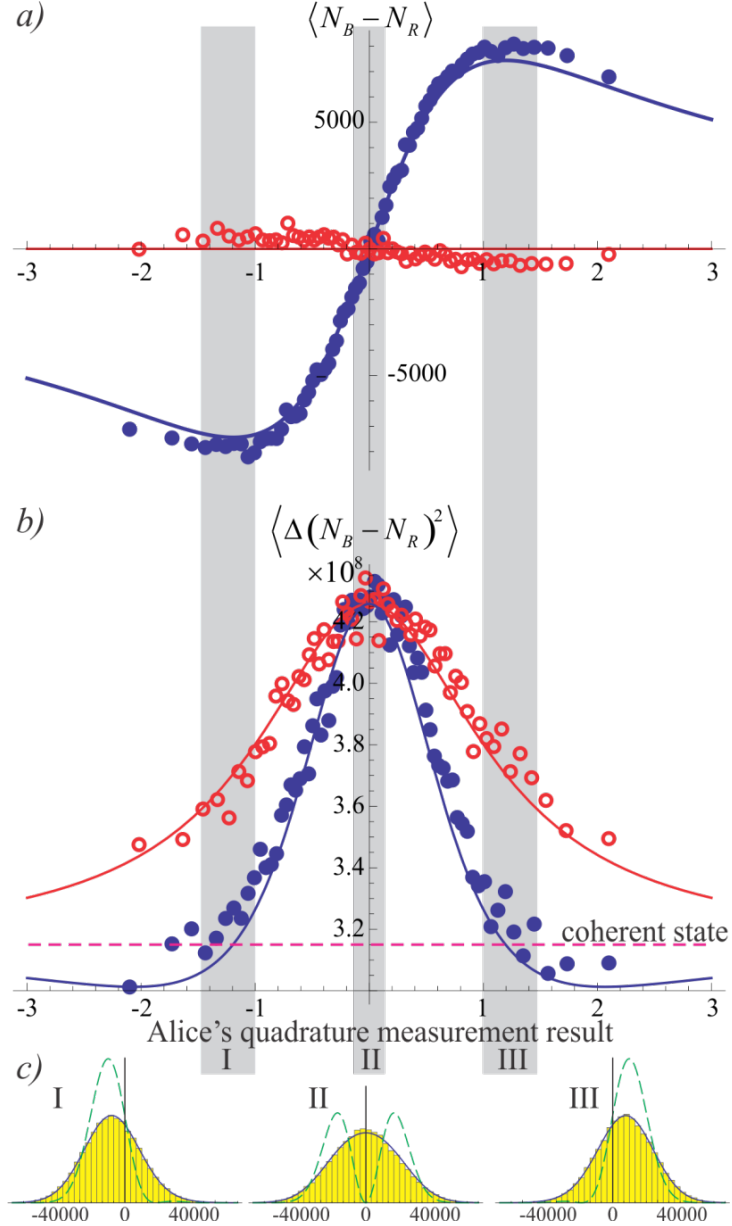

Figure 2: Photon number statistics of the state in Bob's channel that is conditionally prepared by Alice's quadrature measurement. Mean (a) and variance (b) of the difference $N_B - N_R$ between the photon numbers and the reference beam are shown. Blue (solid) circles correspond to the displacement in Bob's channel along the same quadrature as Alice's measurement; for red (hollow) circles the displacement and measurement are in orthogonal quadratures. The purple dashed line in (b) corresponds to $2\alpha^2$, i.e. the variance that would be observed if Bob's channel contained a coherent state of amplitude $\alpha$. Panel (c) displays histograms of $N_B - N_R$ conditioned on Alice's measurement result within intervals I, II, III shown in (a) and (b) by shaded areas. All histograms correspond to the displacement and measurement in the same quadrature. Solid blue and dashed green lines in (c) show theoretical predictions, respectively with and without taking experimental imperfections into account. The statistics represented by histograms I and III, corresponding approximately to states $1/\sqrt{2}\,\hat{D}(\alpha)[|0\rangle \pm |1\rangle]$, can be distinguished by a single energy measurement with a 68% certainty. They are reminiscent of the 'dead' and 'alive' states of Schrödinger's cat.

An interesting interpretation of our results arises if one rewrites state (2) in the superposition basis:

$$|\Psi_D\rangle = \frac{1}{2\sqrt{2}}\left[(|0\rangle+|1\rangle)_A \otimes \hat{D}(\alpha)(|0\rangle+|1\rangle)_B - (|0\rangle-|1\rangle)_A \otimes \hat{D}(\alpha)(|0\rangle-|1\rangle)_B\right]. \quad (4)$$

This, again, can be viewed as Schrödinger's cat, but now the macroscopic terms $\hat{D}(\alpha)(|0\rangle\pm|1\rangle)_B$ have photon number statistics with different mean values of $\alpha^2 + \frac{1}{2} \pm \alpha$ and standard deviations of $\alpha\sqrt{2}$. Performing a single measurement of the photon number observable and checking whether the result exceeds $\alpha^2$ allows one to distinguish these states from each other with an error probability of 10.1%. In other words, the two macroscopic components of our state are distinguishable by means of a single-shot measurement using a detector without microscopic sensitivity.

This observation, which further emphasizes the Schrödinger's cat nature of our state, is confirmed by the experimental results. Alice's observation of quadrature values $X_A$ such that $\psi_0(X_A) = \pm\psi_1(X_A)$ leads, according to Eq. (3), to projecting Bob's channel onto states $\hat{D}(\alpha)(|0\rangle\pm|1\rangle)$. The relevant experimentally observed statistics of Bob's photon number measurement [shown in panels I and III of Fig. 3(c)] are substantially different, albeit not as much as expected theoretically in the idealized setting. This is due to the measurement imperfections discussed above, which increase the probability of error in distinguishing the two states to about 32%.

For direct verification of entanglement, we apply the inverse displacement $\hat{D}(-\alpha)$ to Bob's mode of state (2). Both modes of that state are then subjected to balanced homodyne detection at various local oscillator phases [Fig. 1(c)] [20]. The data output by Bob's homodyne detector exhibit residual phase-dependent quadrature displacement on a scale of $\alpha_r \sim 10$, which we suppress by means of electronic filters. The collected quadrature data are used to reconstruct the density matrix of the two-mode state. This density matrix (Fig. 3) is consistent with a mixture of state (1) with weight η and vacuum state with weight 1−η and shows a high degree of entanglement[27]. Because undisplacement is a local operation, entanglement of the reconstructed state after the undisplacement proves that the micro-macro state was entangled as well.

Finally, we verified robustness of entanglement of state (2) with respect to losses. We inserted a series of attenuators between the displacement and undisplacement operations and reconstructed the density matrix of the resulting state. Fig. 3b shows that, although entanglement is degraded with loss, the rate of this degradation is similar to that expected in the absence of displacement.

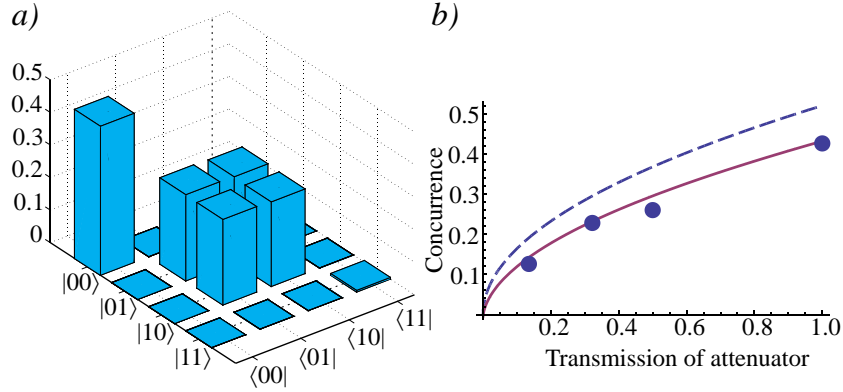

Figure 3: Homodyne tomography of the micro-macro entangled state after "undisplacing" Bob's mode. (a) Density matrix showing entanglement of Alice's and Bob's modes. The plot shows the matrix elements corresponding to zero- and one-photon domains of the optical Hilbert space; the diagonal element contribution from other domains does not exceed 1.2%. (b) Concurrence $C(\hat{\rho}) = 2\left(|\rho_{01}| - \sqrt{\rho_{00}\rho_{11}}\right)$ of the 2-mode state[9] as a function of the attenuation between the displacement and undisplacement operation shows that micro-macro entanglement is robust to optical losses. The dashed theoretical curve corresponds to state (1) and accounts for the losses; the solid curve additionally accounts for the 2-photon term of weight 1.5% that contaminates the heralded single photon.

To summarize, we have demonstrated, for the first time, conclusive experimental evidence of an optical entangled state consisting of two terms that are both macroscopic in the particle number and macroscopically distinct from each other. These features distinguish our work form previous experiments aimed at generating large-size optical coherent superpositions[28,29,30].

We emphasize the difference between our experiment and homodyne tomography of optical states[21]. While the latter also involves interference of a microscopic optical state with a strong field, the fields generated by this interference are viewed as a part of the measurement process – akin to the electronic cascade within an avalanche photodetector. The present work, in contrast, studies these fields as part of a quantum system and unveils their macroscopic and entangled character.

A state similar to ours can be implemented in other quantum systems, for example, in atomic ensembles. A delocalized coherent spin excitation stored in two atomic clouds[31] can be subjected to phase-space displacement by briefly applying a magnetic field perpendicular to the quantization axis, leading to precession of the macroscopic Bloch vector by a small angle. The resulting atomic collective state can be measured and its entanglement verified using the technique of off-resonant Faraday interaction[32].

Our study contributes to the ongoing discussion in the literature regarding the definition of macroscopic quantum superpositions. We have here adopted the most basic definition, a superposition of two states that have macroscopically different expectation values for some physical observable[16]. We have shown that our state is compliant not only with this definition, but with an even stronger criterion: its two components are largely distinguishable by means of single-shot measurements with a macroscopic detector. An additional argument for the macroscopicity of our state is its high sensitivity to certain types of decoherence[19]. However, there are also definitions of macroscopic quantum superpositions that are more stringent, and would exclude the present state[33,34]. We hope that our work will stimulate further investigation and discussion on this topic, which should eventually bring about a much more precise understanding of what we mean by macroscopic quantum effects. In particular, this may lead to a more detailed taxonomy of different 'Schrödinger cats'.

There are more practical questions as well. Although the two terms comprised in state (2) are macroscopically distinct, their difference scales as a square root of their size. This feature is related to the robustness to loss exhibited by our state. Will it be possible to experimentally demonstrate macroscopic entanglement for a state that contains terms whose difference in photon number is comparable to their magnitude? What is the general class of macroscopic entangled optical states that are robust to losses? Will such states be useful for quantum technology, e.g. quantum metrology? Some of these questions are already being discussed in the literature[8,14,35], but more research is required before complete answers are found.

**Acknowledgments.** The work was sponsored by NSERC, CIFAR, AITF. We thank N. Brunner, N. Sangouard, N. Gisin, R. Thew, S. Rahimi-Keshari, S. Raesi, B. Sanders for helpful discussions.

**Methods.** We use a mode-locked Ti:Sapphire laser (Coherent Mira 900) to produce transform-limited pulses of ~1.6 ps width at ~790 nm wavelength and a repetition rate of 76 MHz. The light from this laser is frequency doubled in a single pass through a 17-mm long lithium triborate crystal and subjected to spatial filtering, yielding ~45 mW average power at 390 nm. This field is focused, with a waist of 100 μm, into a 2-mm long periodically poled potassisum-titanyl phosphate crystal for parametric down-conversion, in a type II spatially and spectrally degenerate configuration. The signal and idler photons are separated using a polarizing beam splitter. Idler photons are filtered spatially with a single-mode fiber and spectrally with a 0.3 nm interference filter, and subsequently registered by a PerkinElmer SPCM-AQR-14-FC single-photon detector. Count events occur at a rate of 50-60 kHz. Each such event heralds preparation of a single photon in the signal channel, in a highly pure spatial and spectral mode; however, the signal state features a small (~1.5%) two-photon fraction due to high amplitude of parametric down-conversion.[36]

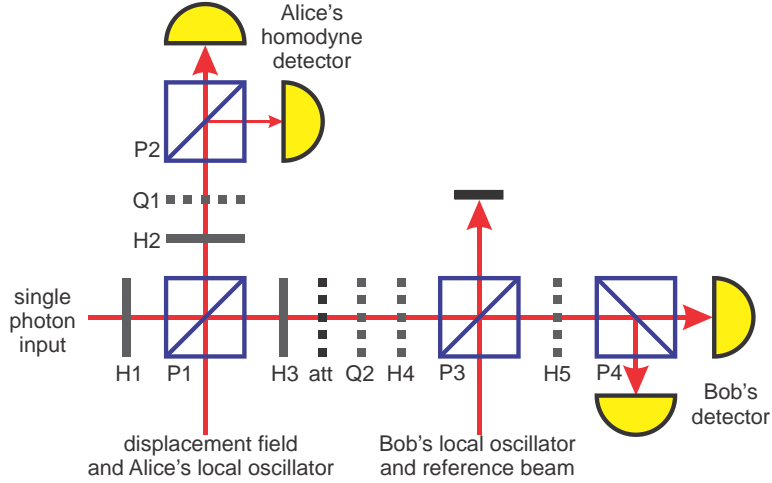

Figure 4: Implementation of the setup. Half-wave plates are denoted by H, quarter-wave plates by Q, polarizing beam splitters by P, attenuator by att. See Methods for further details.

The heralded single-photon is directed into the circuit shown in Fig. 4. It is first split between Alice's and Bob's stations using a half-wave plate H1 and polarizing beam splitter P1. A strong field from the laser is entering the other input port of P1, its horizontally polarized transmitted portion (~10 mW) serving as the local oscillator for Alice's homodyne detector and its vertically polarized reflected portion (~120 mW) providing the displacement field for Bob's channel. Waveplate H3, whose optical axis is rotated by 4.5° with respect to horizontal, mixes the displacement field with Bob's portion of the entangled single photon in the horizontal mode, creating phase-space displacement in that mode[23]. The resulting displaced field power is 3 mW, or $\alpha^2 \approx 1.6 \times 10^8$ photons.

Alice's portion of the single photon is mixed with the local oscillator using H2 and P2 for homodyne detection. In order to change the phase relation between the quadrature measured by Alice and the phase-space displacement of Bob's mode, a quarter-wave plate Q1 is inserted into Alice's channel.

In order to implement the configuration shown in Fig. 1(c), we remove waveplates Q2, H4 and H5. In this way, the displaced mode is transmitted through P3 and P4. Bob's reference beam, on the other hand, is reflected from P3 and P4. For quadrature measurements in Bob's channel [Fig. 1(b)], we insert waveplates Q2 and H4 in order to undo the displacement in the horizontal mode and waveplate H5 to mix the local oscillator and the signal field. In this way, the same balanced detector can be used for both the energy and quadrature (homodyne) measurements at Bob's station. Note that phase locking between the local oscillators in Alice's and Bob's channels was not necessary because the phase drift of these two fields was much slower than the data acquisition rate.